\documentclass[a4paper,11pt]{article}
\usepackage{pos}
\usepackage{wrapfig}
\newcommand{\vx}{\vec{x}}
\newcommand{\vp}{\vec{p}}
\newcommand{\vk}{\vec{k}}

\title{Universality of sound modes in kinetic theory}

\author*[a]{Xiaojian Du}
\author[b]{Stephan Ochsenfeld}
\author[b]{Sören Schlichting}

\affiliation[a]{Instituto Galego de Física de Altas Enerxías (IGFAE), Universidade de Santiago de Compostela,\\
 E-15782 Galicia, Spain}

\affiliation[b]{Fakultät für Physik, Universität Bielefeld,\\
D-33615 Bielefeld, Germany}

\emailAdd{xiaojian.du@usc.es}
\emailAdd{s.ochsenfeld@uni-bielefeld.de}
\emailAdd{sschlichting@physik.uni-bielefeld.de}

\abstract{We present a simple approach to extract hydrodynamic sound modes and non-hydrodynamic modes in kinetic theories from response functions of the energy-momentum tensor. By comparing the response functions in four types of kinetic theories, namely the Relaxation-Time Approximation, scalar $\phi^4$ theory, SU(3) Yang-Mills theory and QCD kinetic theory, we find a remarkable degree of universality for the sound mode, even beyond the hydrodynamic regime. 
}

\notes{\note{\textbf{Acknowledgment}: We thank Weiyao Ke, Aleksas Mazeliauskas, Guy D. Moore, Ismail Soudi, Bin Wu, and Yi Yin for their valuable discussions. This work is supported by the Deutsche Forschungsgemeinschaft (DFG) under grant CRC-TR 211 “Strong-interaction matter under extreme conditions” project no. 315477589-TRR 211. XD is also supported by Xunta de Galicia (Centro singular de investigacion de Galicia accreditation 2019-2022), European Union ERDF, the “Maria de Maeztu” Units of Excellence program under project CEX2020-001035-M, the Spanish Research State Agency under project PID2020-119632GB-I00, and European Research Council under project ERC-2018-ADG-835105 YoctoLHC. The authors gratefully acknowledge computing time provided by the Paderborn Center for Parallel Computing (PC2).}}

\FullConference{%
  HardProbes2023\\
  26-31 March 2023\\
  Aschaffenburg, Germany}

\begin{document}
\maketitle

\section{Introduction}
Collective phenomena observed in relativistic heavy-ion collisions (HICs) signal the existence of the Quark-Gluon Plasma (QGP), a deconfined state of nuclear matter described by the underlying theory of Quantum Chromo Dynamics (QCD). This behavior can be described by relativistic viscous hydrodynamics, an effective theory at a macroscopic level.
Modern constructions of hydrodynamic theories rely on gradient expansions of the energy-momentum tensor of the fluid, which limits the validity of the hydrodynamic description to be only close to equilibrium. Conversely, any microscopic non-equilibrium description is expected to reach its macroscopic hydrodynamic effective description in the long-wavelength and low-frequency limit~\cite{Romatschke:2015gic,Moore:2018mma,Kurkela:2017xis} if the system is sufficiently close to equilibrium.
Constructing effective hydrodynamic descriptions by integrating out the microscopic degrees of freedom may wash out the diversity in microscopic theories. It is thus interesting to study the emergence of macroscopic behavior from different underlying microscopic theories, and to what extent the universal features emerge in the macroscopic limit.

Non-hydrodynamic modes describe the remnants beyond the near-equilibrium hydrodynamic modes~\cite{Romatschke:2017vte}; these may exhibit some degree of universality or display a strong diversity among different microscopic theories out-of-equilibrium.
We investigate these questions within various microscopic kinetic theories~\cite{Arnold:2002zm}, where the microscopic dynamics are governed by the single particle distributions, and infer the macroscopic behavior of the energy-momentum tensor. Based on this description, we present a simple approach to extract the hydrodynamic sound modes and non-hydrodynamic modes. A remarkable degree of universality of the macroscopic behavior is observed, in particular for the hydrodynamic sound mode. Details of our study  are provided in~\cite{Du:2023bwi}.

\section{Linearized kinetic theory}
We study four types of kinetic theories, including the Relaxation Time Approximation (RTA), scalar $\phi^4$ theory (SCL) ~\cite{Mullins:2022fbx}, SU(3) Yang-Mills (YM) kinetic theory~\cite{Kurkela:2015qoa} and QCD kinetic theory~\cite{Kurkela:2018oqw,Du:2020dvp}, in four sets of Boltzmann equations respectively
\begin{eqnarray}
p^\mu\partial_\mu f(t,\vx,\vp)&=&C[f(t,\vx,\vp)].
\end{eqnarray}
Considering linearized perturbations $\delta f$ on top of a homogeneous equilibrium background $f_{\rm eq}$ in the form of $f(t,\vx,\vp)=f_{\rm eq}(T,\vp)+\delta f(t,\vx,\vp)$, a Fourier transform from the position $\vx$ to the wave-number $\vk$
\begin{eqnarray}
\delta f(t,\vk,\vp)=\int\frac{d^3x}{(2\pi)^3}e^{-i\vk\cdot \vx}\delta f(t,\vx,\vp)\;,
\end{eqnarray}
results in a complex integro-differential equation 
\begin{eqnarray}
(p^0\partial_t+i\vp\cdot\vk)\delta f(t,\vk,\vp)&=&\delta C[f_{\rm eq}(T,p),\delta f(t,\vk,\vp)]\;.
\label{eq-linear}
\end{eqnarray}
for the phase-space distributions $\delta f$ of linearized perturbations.
The difference among kinetic theories is encoded in the collision integrals $\delta C[f,\delta f]$.
We source scalar perturbations by small changes in the background temperature $\delta T$, leading to the initial condition
\begin{eqnarray}
\delta f(t=0,\vk,\vp)=\delta T\frac{\partial f_{\rm eq}(T,\vp)}{\partial T}\;.
\end{eqnarray}

\section{Universality of energy-momentum response}
The nontrivial response to a temperature perturbation can be quantified in kinetic theory by means of a response function $G^{s}_{s}$ in the sound channel, which we define in terms of perturbations of energy density $\delta e$ as
\begin{eqnarray}
\tilde{G}^{s}_{s}(\vk,t)=\frac{\delta e(\vk,t)}
{\delta e(\vec{k},t=0)}\;,~~~{\rm with}~~~\delta e(\vec{k},t)=\delta T^{00}(\vec{k},t)=\int\frac{d^3p}{(2\pi)^3}p^0\delta f(t,\vk,\vp)\;.
\label{eq-response}
\end{eqnarray}
A simple example of universality in the hydrodynamic theory can be seen by reformulating the response function of the first-order hydrodynamic theory. Indeed, the response function can be expressed in terms of the scaling variables $\bar{t}=t\frac{sT}{\eta}$, $\bar{k}=k\frac{\eta}{sT}$ as
\begin{eqnarray}
\tilde{G}_{\rm hydro}^{\rm 1st}(k,t)
=\cos(c_skt)e^{-\frac{2}{3}\frac{\eta}{sT}k^2 t}
=\cos(c_s\bar{k}\bar{t})e^{-\frac{2}{3}\bar{k}^2 \bar{t}}\;.
\label{eq-hydro-response}
\end{eqnarray}
The comparison of response functions between kinetic theories for different wave numbers $\bar{k}$ is naturally the first examination of the universal scaling behavior beyond hydrodynamics. 

\begin{figure}[t!]
    \centering
    \includegraphics[width=1.00\textwidth]{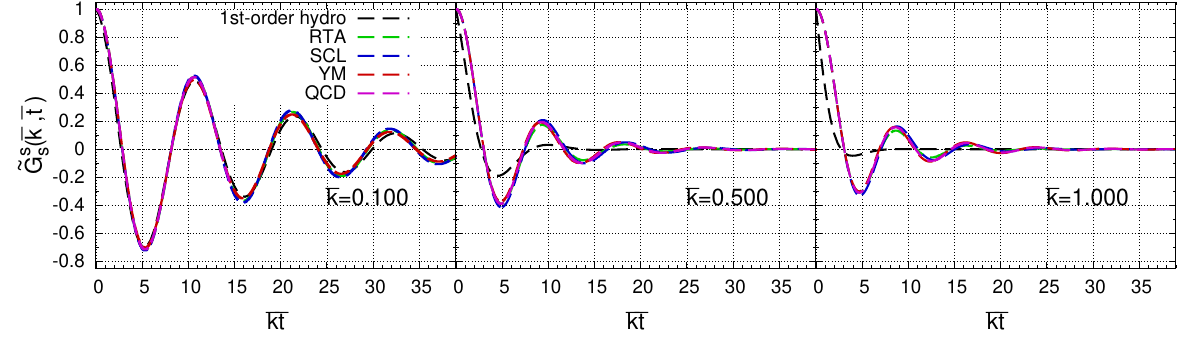}
    \caption{Energy response functions $\tilde{G}_s^s(\bar{k},\bar{t})$ in different kinetic theories (QCD, YM, SCL, RTA) and wave-numbers $\bar{k}$, compared to the first-order hydrodynamic response function in Eq.~(\ref{eq-hydro-response}).
    \label{fig:Response}
    }
\end{figure}

\textbf{Universality of response functions in kinetic theories:} 
We present the comparison of response functions among different kinetic theories as well as the response function from first-order hydrodynamic theory in Fig.~\ref{fig:Response} at various $\bar{k}$.
The left panel of Fig.~\ref{fig:Response} shows a clear convergence of all kinetic theories to the hydrodynamic limit for small gradients $\bar{k}=0.1$. In the middle and right panel, we can see a remarkable degree of universality between kinetic theories for larger wave number $\bar{k}=0.5, 1.0$, well beyond the hydrodynamic limit.

\textbf{Sound and non-hydrodynamic modes:} 
In order to further quantify this universal energy-momentum response, we present a simple approach to extract the hydrodynamic sound modes and the non-hydrodynamic modes. Inspired by the first-order hydrodynamic response function in Eq.~(\ref{eq-hydro-response}), we fit the response functions from kinetic theory in two components
\begin{eqnarray}
\tilde{G}_s^s(\bar{k},\bar{t})=\tilde{G}_s(\bar{k},\bar{t}>4\pi)+\tilde{G}_n(\bar{k}>0.4\pi,\bar{t}),
\end{eqnarray}
with the same functional form containing complex frequencies
\begin{eqnarray}
&\tilde{G}_{s/n}(\bar{k},\bar{t})= Z_{s/n}(\bar{k})\cos[{\rm Re}(\bar{\omega}_{s/n}(\bar{k}))\bar{t}+\phi_{s/n}(\bar{k})]e^{{\rm Im}(\bar{\omega}_{s/n}(\bar{k}))\bar{t}}.
\label{eq:modes}
\end{eqnarray}
\begin{figure}[t!]
    \centering
    \includegraphics[width=1.00\textwidth]{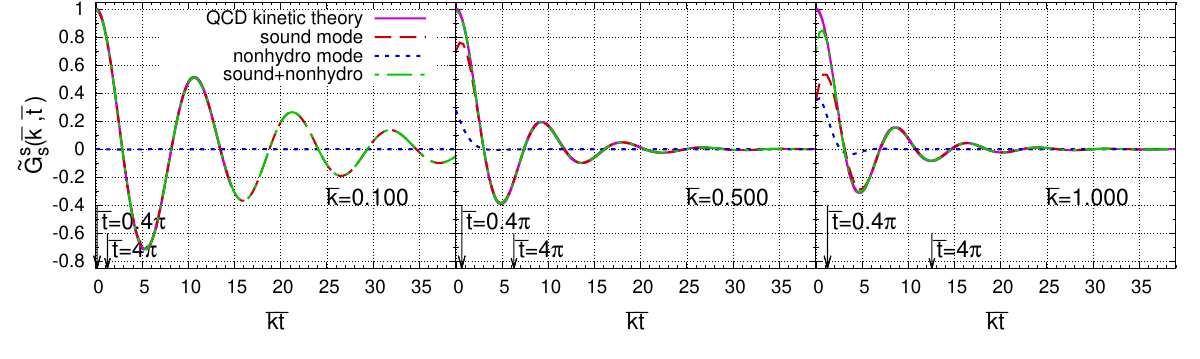}
    \caption{Sound and non-hydrodynamic modes extracted from response functions $\tilde{G}_s^s(\bar{k},\bar{t})$ in QCD kinetic theory at various $\bar{k}$.
    \label{fig:Fit}
    }
\end{figure}
The fitting of the sound mode $\tilde{G}_s$ is done with late time data after the hydrodynamization time $\bar{t}$=$4\pi$, whereas the fitting of the non-hydrodynamic mode $\tilde{G}_n$ is performed for the remaining part on early time scales. 

\begin{wrapfigure}{l}{0.45\textwidth}
    \centering
    \includegraphics[width=0.45\textwidth]{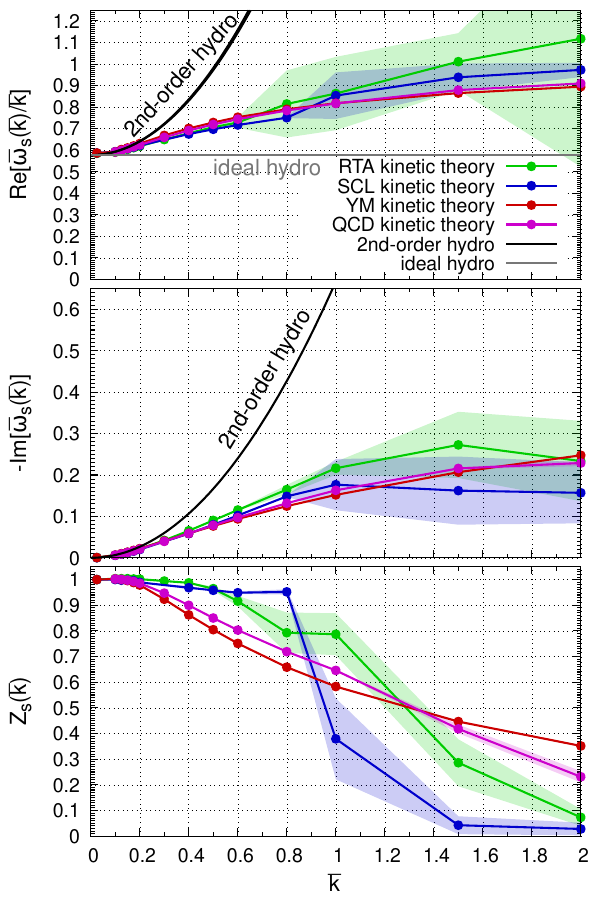}
    \caption{Dispersion ${\rm Re}[\bar{\omega}(\bar{k})]$, damping ${\rm Im}[\bar{\omega}(\bar{k})]$ and residue $Z(\bar{k})$ in different kinetic theories (QCD, YM, SCL, RTA) and wave-numbers $\bar{k}$. Compared to the ideal and second-order hydrodynamic dispersion relations.
    \label{fig:Dispersion}
    }
\end{wrapfigure} 

Since we are using one single mode to describe the non-hydrodynamic behavior, which in general can be expected to exhibit multiple poles or branch-cuts in the complex frequency plane, we avoid fitting very early times before $\bar{t}$=$0.4\pi$. We demonstrate the fitting for the QCD response function in Fig.~\ref{fig:Fit}. The left panel at small $\bar{k}=0.1$ shows the sound mode dominance at the hydrodynamic regime. The non-hydrodynamic mode appears at slightly larger $\bar{k}=0.5$ in the middle panel when the plasma is out of the hydrodynamic regime. The right panel shows an increase in the non-hydrodynamic contribution at even larger $\bar{k}=1.0$ and a failure of fitting at a very early time $\bar{t}<0.4\pi$, where additional non-hydrodynamic excitations play a prominent role.

\textbf{Dispersion, damping and residue:}
After extracting the sound and non-hydrodynamic modes from Eq.~(\ref{eq:modes}), we also inspect the universality in terms of complex frequencies and residues of the hydrodynamic sound modes presented in Fig.~\ref{fig:Dispersion}.
The real part of the frequency represents the dispersion relation of the perturbation and the imaginary part of the frequency represents the damping of the modes. Both dispersion and damping show universality among kinetic theories even at large $\bar{k}$ beyond the hydrodynamic regime. At small $\bar{k}$ they all converge to the hydrodynamic limit as compared to the ideal and second-order hydrodynamic dispersion relations.
The residue characterizes the fraction of energy carried by sound and non-hydrodynamic modes respectively. It goes from sound mode dominance at small $\bar{k}$ down to non-hydrodynamic mode dominance at larger $\bar{k}$ and also shows some level of universality between different microscopic theories.

\begin{figure}[t!]
    \centering
    \includegraphics[width=1.00\textwidth]{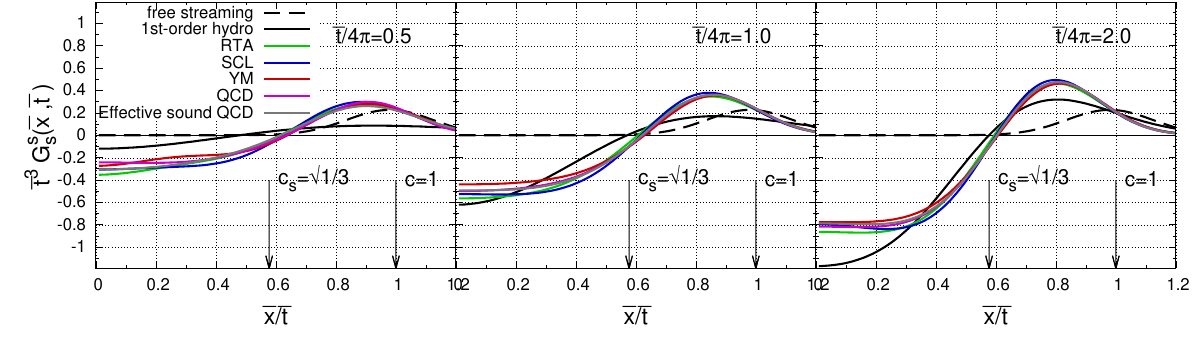}
    \caption{Energy response functions in position space $G_s^s(\bar{k},\bar{t})$ in different kinetic theories (QCD, YM, SCL, RTA) compared to the first-order hydrodynamics as well as free-streaming.}
    \label{fig:Position}
\end{figure}

\textbf{Response functions in position space:}
By Fourier transforming back from wave-number $\bar{k}$ space to the position $x$ space, we evaluate the response function as a space-time propagator
\begin{eqnarray}
G_{s}^{s}(\vec{x},t)=\int\frac{d^3\bar{k}}{(2\pi)^3}e^{i\vk\cdot\vx}\tilde{G}_{s}^{s}(\vk,t)e^{-\sigma (\bar{k}\bar{t})^2} \ ,
\label{eq:fourierposition}
\end{eqnarray}
We show the corresponding response function in position space in Fig.~\ref{fig:Position}.
The universality among all kinetic theories is clearly visible, even for early time, while they approach the hydrodynamic limit at a later time indicating the thermalization of the medium.

\section{Conclusions \& Outlook}
We study the energy response functions of four types of kinetic theories including RTA, SCL, YM, and the QCD. A simple method is performed to extract the sound and non-hydrodynamic modes from the kinetic response functions. The sound modes extracted from the kinetic theories all converge into their universal hydrodynamic limit. 
Among all these studies, we find a remarkable degree of universality among all kinetic theories, especially for sound modes even at large gradients beyond the hydrodynamic regime. 
This indicates the possibility to construct an effective hydrodynamic theory beyond the traditional hydrodynamic validity regime in the future.

\end{document}